\documentclass[a4paper,fleqn]{cas-dc}

\usepackage[numbers]{natbib}

\def\tsc#1{\csdef{#1}{\textsc{\lowercase{#1}}\xspace}}
\tsc{WGM}
\tsc{QE}

\begin{document}
\let\WriteBookmarks\relax
\def\floatpagepagefraction{1}
\def\textpagefraction{.001}
\let\printorcid\relax 

\shorttitle{Multi-Angle Rotational Actuation in a 0.8-mm-Thick Preload-Free Piezoelectric Micromotor}    

\shortauthors{Zhengbao Yang et al.}

\title[mode = title]{Multi-Angle Rotational Actuation in a 0.8-mm-Thick Preload-Free Piezoelectric Micromotor}

\author[1]{Haijia Yu}

\author[1]{Mingtong Chen}

\author[1]{Zhengbao Yang}
\cormark[1] 
\ead{zbyang@hk.ust} 
\ead[URL]{https://yanglab.hkust.edu.hk/}

\address[1]{The Hong Kong University of Science and Technology
Hong Kong, SAR 999077, China}

\cortext[1]{Corresponding author} 

\begin{abstract}
Micro motors can be used in numerous fields like Micro medical testing and treatment. To achieve a smaller size, micro piezoelectric motors in laboratories often omit the outer casing, which can lead to functional defects such as rotation only in one fixed direction or the need for external weights (which are not counted within the motor's volume) to increase preload. However, this significantly reduces the practical value of micro piezoelectric motors. This paper proposes a new driving principle for piezoelectric motors to design a micro piezoelectric motor that can rotate at a wide range of angles (e.g. up to ±80°)without increasing the motor's casing and does not require external weights, with a stator thickness of only 0.8 mm. This motor has significant application potential in OCT endoscopes and thrombectomy grinding heads

\end{abstract}




\maketitle

\section{Introduction}
Micro piezoelectric has great application value in different fields, for example, it can be placed at the end of Optical Coherence Tomography (OCT) endoscopes\cite{1}, shown in figure 1. OCT endoscopes combine fiber optic transmission with endoscopic observation to visualize the condition of deep tissues. This type of endoscope requires a motor installed at the tip to drive the triangular reflector to rotate, enabling circular scanning. Additionally, at present, the size of the arterial vascular calcification plaque grinding head can only be about 2mm \cite{2}, shown in figure 2, and the micro piezoelectric motor can further reduce the size of the ablation head to clear thinner arteries.

\begin{figure*}[h]
	\centering
		\includegraphics[scale=1]{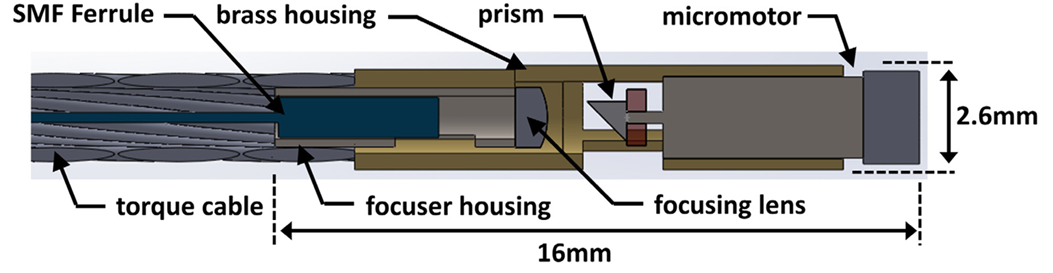}
	  \caption{Structure of OCT endoscope}
      \label{FIG:1}
\end{figure*}

\begin{figure}[h]
	\centering
		\includegraphics[scale=0.6]{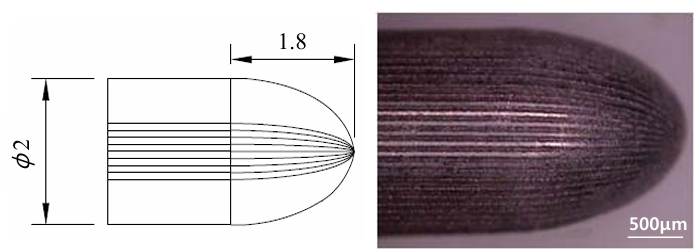}
	  \caption{Arterial vascular calcification plaque ablation head}
      \label{FIG:2}
\end{figure}

Piezoelectric motors (here focusing on rotary piezoelectric motors) utilize the converse piezoelectric effect of piezoelectric sheets bonded to the stator to convert electrical energy into mechanical vibrations. These vibrations are synthesized and transmitted through friction to the rotor, driving it to rotate. Micro piezoelectric motors are generally composed of two parts: an axial component and a hole component, which can also be divided into two types. The first one is the Circular Tube Piezo-motor, which has the hole component as the stator, the axial component as the rotor; The second one is the Cylindrical Piezo-motor has the hole component as the stator, the axial component as the rotor. In 2003, Dong Shuxiang and others developed a piezoelectric motor with a direct diameter of 1.5 mm, which belong to Circular Tube piezo-motor, featuring a resonant frequency of 67 kHz, and a maximum rotational speed and stall torque of 2000 r/min and 500 µNm, respectively\cite{3}.In 2005, Zhou Tieying, Tsinghua University, et al designed the first Cylindrical Piezo-motor\cite{4}, which has a diameter of 1 mm. The motor speed is 1800 r/min and the blocking force reaches 4N·m by using the bending vibration mode.

Advances in micro piezoelectric motors over the next decade were largely based on these two type of  motors.\cite{5,6,7,8} Until 2016, Tomoaki Mashimo designed a micro piezoelectric motor using a cube with a side length of 0.5 mm, which is the smallest piezo-motor in the world.\cite{9}However, this micro piezoelectric motor can only rotate when paired with a fixed fixture, Additionally, two weights need to be added at both ends of the rotor to provide preload between the rotor and the stator (This ultrasonic motor would be much larger if the volume of the fixture and counterweight were taken into account). This makes the motor unsuitable for applications requiring different angles and directions, such as OCT endoscopes and arterial grinding heads that need to navigate through the lumen and blood vessels in a winding manner. This motor cannot function properly.

In 2023 and Liu Zhoulong and others designed a micro piezoelectric motor\cite{10}, removing the outer shell of the motor to reduce its diameter, with a minimum diameter of 1 mm. However, this design provides preload by adding gaskets above the rotor to apply downward pressure, which has the same drawback as mentioned above: it can only rotate in a fixed direction\cite{11,12}.

\section{Method}

In order to overcome the above disadvantages of the motor and make the motor as small as possible, it is necessary to start from the driving principle of the micro piezoelectric motor. There are obvious differences between the driving principle of the circular tube piezoelectric motor and the cylindrical piezoelectric motor. According to the Tomoaki Mashimo’s article\cite{9}, the circular tube piezoelectric motor drive mainly relies on applying different alternating voltages E1 and E2 on two piezoelectric sheets that are vertical to each other.

$$
E_1=Asin(2\pi ft) 
$$
$$
E_2=Asin(2\pi ft-\pi/2)
$$
Among them, E1 can generate a three-wave mode around the stator holes, and E2 can excite another three-wave, as shown in Figure 3. However, considering the effect of the increased preload of the heavy weight, the rotor only has close contact with the lower part of the stator, and the main driving force is also generated in this part, which may lead to the rotation center of the mover and the geometric center of the stator not coinciding, which gives new ideas to the driving principle.

\begin{figure}[h]
	\centering
		\includegraphics[scale=0.6]{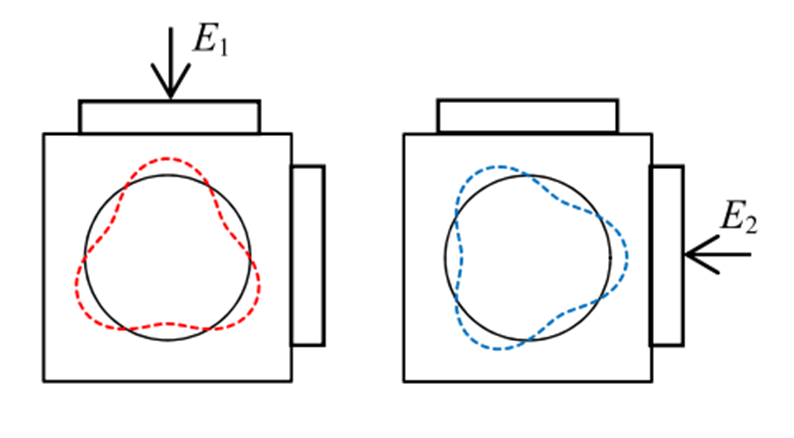}
	  \caption{Drive principle of cylindrical piezoelectric motor}
      \label{FIG:3}
\end{figure}

The centripetal friction drive mode is used as the final driving principle, which requires the inner diameter of the rotor and the output shaft of the stator to be a loose clearance fit. The inner diameter of the rotor is 0.6mm and the outer diameter of the stator output shaft is 0.5mm. 316F Stainless steel is used as raw material.The rotor is designed in the shape of three stepped shafts, with the outermost layer used to reduce the contact area with the stator metal platform. This part employs the drive principle described in Section 2.2. A smaller contact area allows for a greater elliptical trajectory friction distance, resulting in better driving performance. The middle layer is hollowed out to serve two purposes: one is to reduce weight, and the other is to increase the eccentricity when the rotor is placed vertically.

This groove facilitates the engagement with the outer edge of the lower segment of the rotor, generating friction. The stator model is shown in Figure 4. (a), and the actual stator is illustrated in Figure 4. (b). The stator is a cylindrical shape with a diameter of 1mm and a height of 6mm. A rectangular strip measuring 0.6mm by 6mm is removed from the circumference to accommodate the piezoelectric sheet. After this removal, the stator's thickness is 0.8mm.

\begin{figure}[h]
	\centering
		\includegraphics[scale=0.5]{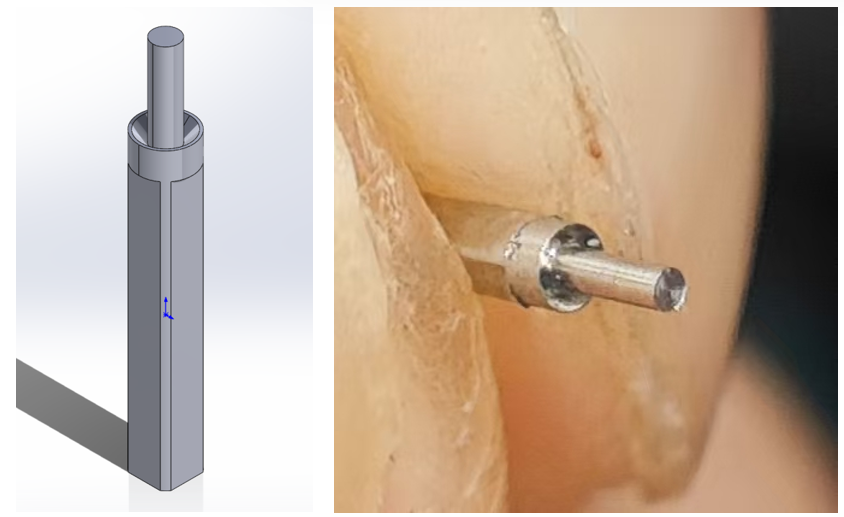}
	  \caption{Stator Model and Actual Stator}
      \label{FIG:4}
\end{figure}

The selection of piezoelectric material is crucial. Here lies an important parameter often overlooked: the mechanical quality factor Qm. Qm determines the efficiency of converting electrical energy into mechanical vibrations. If the mechanical quality factor of the piezoelectric material is too low, most of the electrical energy will be converted into heat, causing the piezoelectric to adhere poorly or the solder to melt. Generally, hard piezoelectric ceramics like PZT8 have a higher mechanical quality factor.However, it is difficult to process the hard piezoelectric sheet. Limited by experimental conditions, the size of the piezoelectric sheet used is a rectangular 5×0.8×0.8mm. The author believes that if a thinner piezoelectric sheet is used, similar experimental results can be achieved. Figure 11 shows the size of the stator, rotor and piezoelectric sheet. It can be seen that the thickness is less than 1mm, about 0.8mm.

Due to the extremely small size of the piezoelectric sheets, they cannot be made into flanging electrodes, so the stator column must serve as the negative electrode of the piezoelectric sheet. This means that the piezoelectric sheet and the stator must be conductive. After testing, it was found that the viscosity of conductive silver paste is insufficient After electrification, they cannot cause the stator column to oscillate. The particles in the nickel-carbon powder mixed adhesive are too large, making it difficult to adhere firmly. Therefore, to meet both conductivity and strong adhesion requirements, a three-stage coating method is used. 

In order to find the driving resonant frequency of the piezoelectric motor, the impedance and phase angle of the piezoelectric sheet are observed by using the impedance meter. The horizontal coordinate of the trough displayed by the impedance meter is the resonant frequency of the piezoelectric sheet.The signal generator is adjusted to apply a voltage of 250V and outputs signals at different frequencies for frequency sweeping. The curve showing the change in vibration speed from 0 to 100KHz is plotted as shown in Figure 5. From the figure, it can be seen that when the frequency is 64.9KHz, the ultrasonic motor stator has the best vibration effect.

\begin{figure}[h]
	\centering
		\includegraphics[scale=0.4]{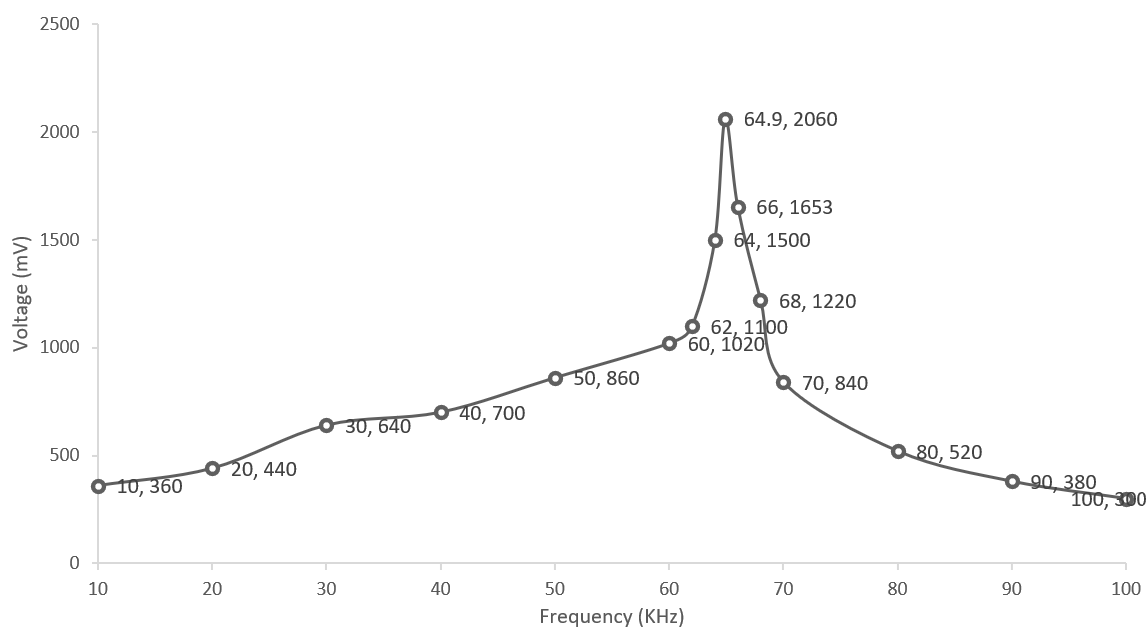}
	  \caption{Scan the Vibration Curve}
      \label{FIG:5}
\end{figure}

To verify the correctness of the driving principle, COMSOL is used for modal simulation and motion trajectory simulation of the piezo-motor. A fixed constraint is applied to the lowest plane of the motor, with the stator column material being stainless steel and the piezoelectric ceramic material being PZT8. A voltage with a phase angle difference of 90 degrees is applied across two vertical piezoelectric sheets.Next, the motion trajectories of any point on the outer wall of the output shaft and any point on the edge of the metal cap during the stator vibration are plotted, with the results displayed in Figures 6.

\begin{figure}[h]
	\centering
		\includegraphics[scale=0.4]{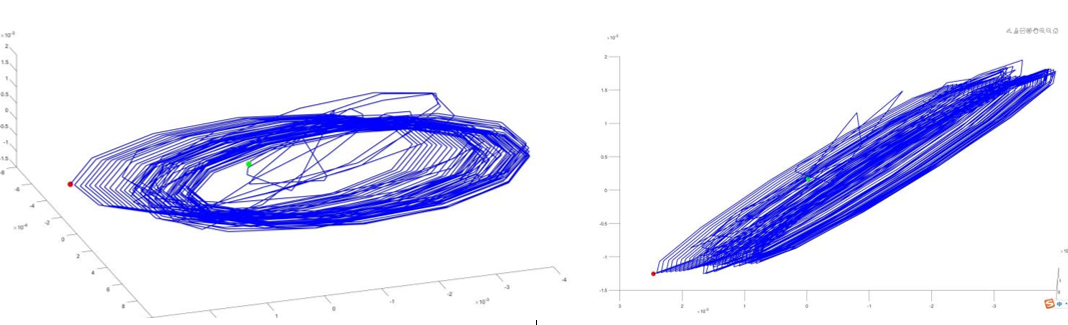}
	  \caption{Trajectory of the Shaft and Trajectory of the Metal Cap}
      \label{FIG:6}
\end{figure}

In summary, voltage, frequency, and inclination angle all influence the rotational speed. The effect of voltage is relatively linear. However, if the voltage drops below a certain threshold, the motor will fail to operate. In such cases, adjusting the drive frequency can further reduce the rotational speed. When the inclination angle approaches ±90 degrees, meaning the motor is placed vertically, the rotational speed becomes extremely unstable. This instability can be mitigated by reducing the eccentricity by decreasing the stator hole diameter. However, this approach also reduces the centripetal force, necessitating further design exploration.

Use a high-speed camera to test the speed of the motor, shown in Figure 7. (a). The high-speed camera can reach 1,000 frames per second and can be used in editing software to observe the number of rotations of the rotor.

The rotational speed is most affected by the voltage. When the phase difference between the two electrodes is 90 degrees, the voltage frequency is 64.9KHz, and the motor is placed horizontally, the voltage change is adjusted to get Figure 7. (b). It can be seen from the figure that the rotational speed will increase with the increase of the voltage, and the highest rotational speed can be reached to 882r/min. The voltage below 100V cannot make the motor rotate stably.

The frequency of the two piezoelectric sheets is adjusted, and the voltage is kept at 200V. The motor is placed horizontally, and the speed and frequency curve are shown in Figure 7 (c). It can be seen from the figure that the lowest speed of the motor is 90r/min.

Keep the voltage and frequency unchanged, and adjust the Angle between the motor and the plane. The relationship between the Angle and the speed is shown in Figure 7 (d). As can be seen from the figure, the speed change is not large within ±80 degrees of inclination. However, after 80 degrees, the motor rotor is very unstable and is likely to stop. This may be related to the aperture and size weight of the rotor.

In summary, voltage, frequency, and inclination angle all influence the rotational speed. The effect of voltage is relatively linear. However, if the voltage drops below a certain threshold, the motor will fail to operate. In such cases, adjusting the drive frequency can further reduce the rotational speed. When the inclination angle approaches ±90 degrees, meaning the motor is placed vertically, the rotational speed becomes extremely unstable. This instability can be mitigated by reducing the eccentricity by decreasing the stator hole diameter. However, this approach also reduces the centripetal force, necessitating further design exploration.The torque of the brake determines how much load the motor can carry at most. Because the motor rotor is too small, the load is measured indirectly by measuring angular acceleration and calculating the moment of inertia.The rotor can be regarded as a whole composed of three annular columns with the same outer diameter and different inner diameters. The moment of inertia of the rotor is calculated below.After observation, the motor can be accelerated to the highest speed in about 0.4s when placed horizontally.

\begin{figure*}[h]
	\centering
		\includegraphics[scale=1]{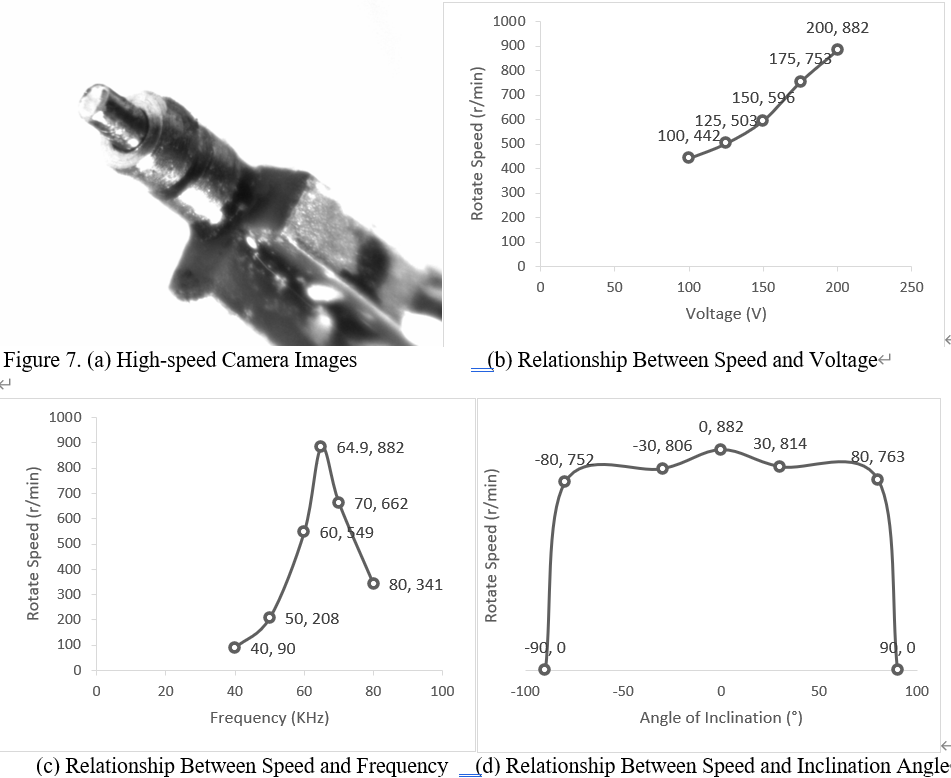}
	  \caption{Motor Speed Test}
      \label{FIG:7}
\end{figure*}

\begin{figure*}[h]
	\centering
		\includegraphics[scale=1]{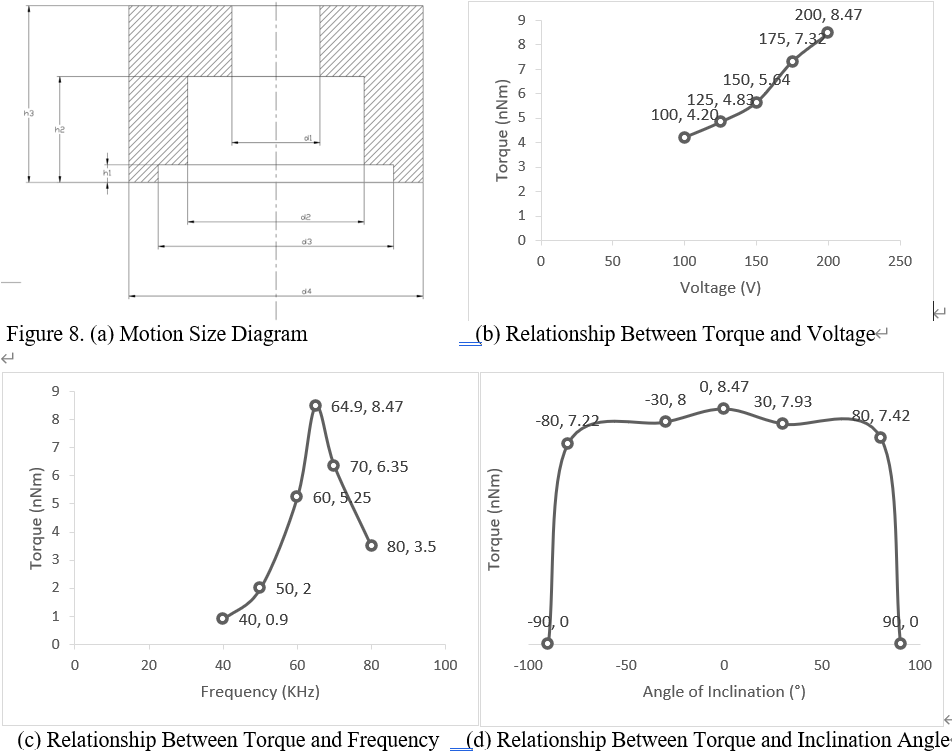}
	  \caption{Rotational Toque Test}
      \label{FIG:8}
\end{figure*}

\section{Summary}

In the first part of this paper, the specific applications of micro piezoelectric motors in the medical field are introduced, such as OCT endoscope and arterial thrombus grinding head. The development history of micro piezoelectric motors is studied, and the piezoelectric motors are divided into two categories: Circular tube motors and cylindrical motors. The second part introduces the driving principles of both cylindrical and tubular motors. Inspired by the eccentric rotation of tubular motors, a third driving method for piezoelectric motors: Centripetal Friction is proposed. This method uses centrifugal force to apply preload, eliminating the need for a motor housing and further reducing the size of the piezoelectric motor. Based on the principles outlined in the second part, the third part designs the motor structure to meet the eccentricity requirements. The fourth part details the manufacturing process of the motor. The fifth part describes the experimental procedures, including verifying the piezoelectric properties using an impedance meter, identifying the resonant frequency through a laser vibrometer sweep, and conducting power-on experiments under simulated conditions that meet the assumptions, to measure the motor's speed and torque under various operating conditions.

This paper introduces an innovative principle-based design that enables the motor to rotate at various angles without requiring additional preload. However, there are still several areas for improvement. First, the stator can be further reduced in size and use thinner piezoelectric sheets. Second, the impact of the clearance fit between the stator and rotor on the motor's speed and torque can be further explored. Third, since some grinding heads are eccentric during thrombus removal, this could potentially enhance the motor's efficiency, though further research is needed. Fourth, integrating the centrifugal friction drive principle with the cylindrical motor drive principle could ensure good driving performance even when the motor approaches ±90 degrees. Fifth, exploring ways to increase the motor's robustness to prevent the rotor from flying off the output shaft.










\bibliographystyle{cas-model2-names}

\bibliography{cas-refs}



\end{document}